\documentclass[preprint,showpacs,preprintnumbers,amsmath,amssymb,nofootinbib]{revtex4}

\usepackage{graphicx}
\usepackage{color}
\usepackage{dcolumn}
\usepackage{bm}

\begin{document}

\preprint{APS/123-QED}

\title{Cosmological perturbations in extended electromagnetism. General gauge invariant 
approach}

\author{Roberto Dale}
\email{rdale@dfists.ua.es}  
\affiliation{%
Departamento de F\'{\i}sica, Ingenier\'{\i}a de Sistemas y Teor\'{\i}a de la Se\~nal,
Universidad de Alicante, 03690, San Vicente del Raspeig, Alicante, Spain    \\
}%

\author{Diego S\'aez}
\email{diego.saez@uv.es}
\affiliation{Departamento de Astronom\'{\i}a y Astrof\'{\i}sica, Universidad de Valencia,
46100 Burjassot, Valencia, Spain\\
}%

\date{\today}

\begin{abstract}
A certain vector-tensor (VT) theory is revisited. It was proposed  
and analyzed as a theory of electromagnetism without the standard gauge invariance. 
Our attention is first focused on a detailed variational formulation of the theory,
which leads to both a modified Lorentz force and the true energy momentum tensor
of the vector field.
The theory is then applied to cosmology.  A complete gauge invariant treatment 
of the scalar perturbations is 
presented. For appropriate gauge invariant variables describing the scalar modes
of the vector field (A-modes), it is proved that  
the evolution equations of these modes do not involve
the scalar modes appearing in General Relativity (GR-modes),
which are associated to the metric and the energy momentum tensor of 
the cosmological fluids. However, the A-modes
modify the 
standard gauge invariant equations describing the GR-modes. 
By using the new formalism, the evolution equations of the A-perturbations are 
derived and separately solved and, then, the correction terms 
--due to the A-perturbations-- appearing in 
the evolution equations of the GR-modes
are estimated. The evolution of these correction terms is studied 
for an appropriate scale. The relevance of these terms   
depends on both the spectra and the values of the normalization constants involved 
in extended electromagnetism.
Further applications of the new formalism will be presented elsewhere.

\end{abstract}



\maketitle

\section{Introduction}
\label{intro}

In previous papers, an extended theory of electromagnetism 
was proposed \cite{bm091} and developed \cite{bm092,bm111,bm112}.
The evolution                                   
of cosmological scalar perturbations was studied in \cite{bm092}; nevertheless, 
the authors assumed that the {\em scalar} perturbations 
of the electromagnetic field do not affect metric perturbations evolution.
They stated that this assumption holds 
both in the radiation dominated period and in the matter dominated era, that is to say, 
at any moment before dark energy domination. Hence, 
it affects the choice of the initial conditions for 
numerical integrations, which are always fixed in the radiation
dominated era ($z \sim 10^{8}$), when the cosmologically significant scales 
had superhorizon sizes. Moreover, in paper \cite{bm092}, the conservation law of standard 
electromagnetism, $\nabla_{\mu} J^{\mu} =0$, is assumed, e.g.,  
to get Eq. (2.3) from Eq. (2.2). Nevertheless, this law is not an equation 
of extended electromagnetism, whose true conservation law (see below) 
admits solutions with $\nabla_{\mu} J^{\mu} \neq 0$. This fact 
is important in cosmology, where the law $\nabla_{\mu} J^{\mu} =0$
implies that only vector modes are involved in the expansion of the current
$J^{\mu} $, whereas the condition
$\nabla_{\mu} J^{\mu} \neq 0$ requires the existence of
$J^{\mu} $ scalar modes. We have introduced one of these modes 
(see below) in our general calculations. 
We see that, in extended electromagnetism, there are scalar modes 
in the expansions of both the electromagnetic field $A^{\mu} $ (A-modes) and 
the current $J^{\mu}$ (J-modes). Since we have not convincing arguments proving that,
initially, at $z \sim 10^{8}$, all these modes are negligible against 
the small scalar modes of the radiation fluid, we cannot ensure that they
do not affect metric perturbations. Hence, the A and J scalar modes
should not be neglected {\em a priori} in order to  
get the metric and fluid initial conditions for numerical integrations.
After these considerations it seems that the assumption used in \cite{bm092}
is a simplifying condition which would
require further justification (if it exists). In this situation, it is obvious that 
a more general study of cosmological perturbations
is worthwhile. It is performed in this paper, where a general complete
treatment of the cosmological perturbations is developed in 
the framework of extended
electromagnetism. Our approach has various relevant properties: 
(i) it is gauge invariant, (ii) it does not 
involve approximating conditions, (iii) it involves a $J^{(0)} $ scalar mode
as it is required by the general conservation law of extended 
electromagnetism, and (iv) it uses appropriate scalar modes 
for the field $A^{\mu } $ which evolve independently of the 
scalar GR-modes (metric and fluid modes).

This paper is structured as follows:
The basic equations of the VT theory are derived --by using variational 
techniques-- in Sec.~\ref{sec-2}, where the energy momentum tensor and
the Lorentz force are calculated.
The theory is applied to cosmology in Sec.~\ref{sec-3},
where the linear approximation is studied by using  
standard techniques based on decoupled scalar, vector, and 
tensor modes \cite{bar80,huw97}. 
An analysis of the equations derived in Sec.~\ref{sec-3} 
is performed in Sec.~\ref{sec-4}, where 
the equations satisfied by the scalar perturbations of 
the vector field are numerically solved, and the 
differences between the equations describing 
the evolution of the scalar modes in General Relativity (GR) and in VT are studied. 
The Appendix contains some estimates in the framework of GR,
which are useful in Sec.~\ref{sec-4}.
Finally, Sec.~\ref{sec-5}
contains a general discussion and our main conclusions.

Let us finish this section fixing some notation criteria. Latin (Greek) 
indexes run from 1 to 3 (0 to 3). The
gravitational constant, the scale factor, the conformal time, and the
Hubble constant are denoted $G$, $a$, $\tau $, and $H_{0} $, 
respectively. Whatever function $D $ may be, $D_{/\alpha}$ stands for 
its partial derivative with respect to the coordinate $x^{\alpha}$, 
and $D_{B}$ represents its background value.
Units are chosen in such a way that the speed of light is $c=1$.
Spatial distances are given in Megaparsecs.

\section{The theory: variational formulation and basic equations}   
\label{sec-2}

A charged isentropic perfect fluid is considered in the framework of the 
VT generalization of Einstein-Maxwell theory 
proposed in \cite{bm091}. The basic equations are derived from the following 
action: 
\begin{equation}
I = \int \left[ \frac {R} {16\pi G}
- \frac {1}{4} F^{\mu \nu } F_{\mu \nu }
+\gamma (\nabla_\mu A^{\mu})^{2} 
+ J^{\mu} A_{\mu} - \rho (1 + \epsilon)
\right] \,\sqrt { - g} \,d^4 x  \ ,
\label{1.1}
\end{equation}   
where $\gamma $ is an arbitrary parameter, 
$R$, $g_{\mu \nu}$, and $g$ are the scalar curvature, the covariant metric components, and
the determinant of the $g_{\mu \nu}$ matrix, respectively. The vector field of the
theory is $A^{\mu} $.
The symbol $\nabla $ stands
for the covariant derivative and we define
$F_{\mu \nu} = \nabla_{\mu} A_{\nu } - \nabla_{\nu} A_{\mu }$. 
The electrical current is $J^{\mu} = \rho_{q} U^{\mu} $, where $\rho_{q} $
is the density of electrical charge and $U^{\mu} $ is the four-velocity of 
the fluid world lines. Finally, for an isentropic perfect fluid, one can 
introduce a conserved energy density $\rho $ [$\nabla_{\mu} (\rho U^{\mu})=0$] and an 
internal energy $\epsilon$; so the fluid energy density is 
$\mu = \rho (1 + \epsilon)$ and the pressure is
$P=\rho^{2}(d\epsilon/d\rho)$ (see \cite{haw99}).

Some VT theories were proposed in the early seventies (see \cite{wil93,wil06}). All these
theories were based on the action:
\begin{eqnarray}
I &=& \left( {16\pi G} \right)^{ - 1} \int {\left( {R + \omega A_\mu  A^\mu  R
+ \eta R_{\mu \nu }
A^\mu  A^\nu   - \varepsilon F_{\mu \nu } F^{\mu \nu } 
+ \tau \,\nabla _\nu  A_\mu  \nabla ^\nu  A^\mu
} + L_{m} \right)} \nonumber \\
& & 
\sqrt { - g} \,d^4 x
\label{1.2}
\end{eqnarray}   
where $\omega$, $\eta $, $\varepsilon$,
and $\tau$ are arbitrary parameters and $L_{m}$ is the matter Lagrangian,
which couples matter with the fields of the VT theory.
Actions (\ref{1.1}) and (\ref{1.2}) are equivalent for 
$\omega =0$, $2\varepsilon - \eta = 8\pi G$, $\tau = \eta = 16\pi G \gamma$,
and $L_{m} = J^{\mu} A_{\mu} - \rho (1 + \epsilon)$.

According to the variational techniques described in \cite{haw99}, three fields 
may be independently varied in the action (\ref{1.1}). These fields are: the vector field 
of the theory ($A_{\mu}$), the flow lines of the fluid ($U^{\mu} $), and the metric field
($g_{\mu \nu}$).

We first vary the field $A_{\mu}$ for fixed flow lines and metric
($\delta_{_{A}}$ variations). Thus, we easily 
obtain the field equations for the $A^{\mu}$ field, whose form is:
\begin{equation}
\nabla^{\nu} F_{\mu \nu} = J_{\mu} + J^{^{A}}_{\mu} \ ,
\label{1.3}
\end{equation} 
where $J^{^{A}}_{\mu} = -2 \gamma \nabla_{\mu} (\nabla \cdot A)$ 
with  $\nabla \cdot A = \nabla_{\mu} A^{\mu} $. Then, from these field equations one easily 
gets the relation: 
\begin{equation}
\nabla^{\mu} J_{\mu} = - \nabla^{\mu} J^{^{A}}_{\mu}   \ ,
\label{1.4}
\end{equation}
which indicates that the total current $J_{\mu}+J^{^{A}}_{\mu}$ is conserved
in the theory. This is the conserved current associated to the  
invariance of action (\ref{1.1}) under the residual gauge transformation ${A}^{\prime}_{\mu} =
A_{\mu} + \partial_{\mu} \phi $ with $\partial^{\mu} \partial_{\mu} \phi = 0$.

In a second step, 
the flow lines are varied for fixed $A_{\mu} $ and $g_{\mu \nu}$ ($\delta_{_{U}}$
variations) and, moreover,                                       
the densities $\rho $ and $\rho_{q} $ are adjusted to satisfy the equation
$\nabla_{\mu}(\rho U^{\mu})=0$ and Eq.~(\ref{1.4}), respectively (see \cite{haw99}). 
Since the right hand side of Eq.~(\ref{1.4}) does not depends on $U^{\mu}$,
the following relation is satisfied $\delta_{_{U}} (\nabla_{\mu} J^{\mu}) 
= \nabla_{\mu} (\delta_{_{U}} J^{\mu}) = 0$. On account of these considerations,
the following equations are easily obtained \cite{haw99}:
\begin{equation}
(\mu + P)U^{\mu}\nabla_{\mu}U^{\nu} = -\nabla_{\mu}P(g^{\mu \nu}+U^{\mu}U^{\nu})+
F^{\mu \nu}J_{\mu}+(\nabla^{\mu}J^{^{A}}_{\mu})A^{\nu} \ .
\label{1.5}   
\end{equation}
These equations describe the fluid evolution in the VT
theory. The last term is the generalized Lorentz force, $f^{^{L}}$, of the 
theory; hence, we can write:              
\begin{equation}
f^{^{L}}_{\nu} = F_{\nu \mu}J^{\mu}+(\nabla^{\mu}J^{^{A}}_{\mu})A_{\nu}
\label{1.6}
\end{equation}

Finally, the metric is varied whereas fields $A_{\mu} $ and $U^{\mu }$ are
fixed ($\delta_{_g}$ variations). For this kind of variations, 
Eq~(\ref{1.4}) leads to the relation 
$\delta_{_g} (\sqrt {-g} J^{\alpha}) = -\delta_{_g} (\sqrt {-g} J^{^{A} \alpha}) $
and, then, from this relation and the 
identity $\nabla^{\mu}[A_{\mu} (\nabla \cdot A)] = (\nabla \cdot A)^{2} 
+ A_{\mu} \nabla^{\mu} (\nabla \cdot A) $, it follows that the Lagrangian 
densities $\gamma (\nabla_\mu A^{\mu})^{2}                             
+ J^{\mu} A_{\mu}$ [involved in action (\ref{1.1})]
and $-\gamma (\nabla_\mu A^{\mu})^{2} $ are fully equivalent.
On account of this fact, $\delta_{_g}$ variations lead to:
\begin{equation}  
G^{\mu \nu} = 8\pi G T^{\mu \nu} \ ,
\label{1.7}
\end{equation}
where $G^{\mu \nu}$ is the Einstein tensor and $T^{\mu \nu}$ is the energy momentum 
tensor of the charged fluid plus the electromagnetic field, whose form is:
\begin{eqnarray}
T^{\mu \nu} &=& (\mu + P)U^{\mu}U^{\nu} + Pg^{\mu \nu} + F^{\mu}_{\,\,\,\, \alpha}F^{\nu \alpha}
- \frac {1}{4} g^{\mu \nu} F_{\alpha \beta} F^{\alpha \beta} \nonumber \\ 
& &
+2\gamma [ \{A^{\alpha}\nabla_{\alpha} (\nabla \cdot A) + \frac {1}{2}(\nabla \cdot A)^{2}\}
g^{\mu \nu}-
A^{\mu}\nabla^{\nu} (\nabla \cdot A) - A^{\nu}\nabla^{\mu} (\nabla \cdot A)
] \ .
\label{1.8}
\end{eqnarray}
This energy momentum tensor is to be compared with that given in \cite{bm092}
taking into account differences in the assumed signature.

Equations~(\ref{1.3}), (\ref{1.5}), and (\ref{1.7}) are the field equations 
of the theory. Any solution of these equations satisfies Eq.~(\ref{1.4}), 
and also the relations $\nabla_{\mu} T^{\mu \nu} = 0$. These last relations
combined with all the field equations and Eq.~(\ref{1.8}) lead to:
\begin{equation}  
\left[U^{\beta}\nabla_{\beta} (\mu) + (\mu + P) \nabla_{\beta} U^{\beta}\right]U^{\alpha} =
2F^{\alpha \beta} J^{^{A}}_{\beta} - 2 A^{\alpha} \nabla^{\beta} J^{^{A}}_{\beta} \ .
\label{1.9}   
\end{equation}
In the Einstein-Maxwell theory, which is formally obtained in the limit $J^{^{A}}_{\beta}
\longrightarrow 0$, Eq~(\ref{1.9}) reduces to 
$U^{\beta}\nabla_{\beta} (\mu) + (\mu + P) \nabla_{\beta} U^{\beta} = 0$ 
(which is identical to Eq.~(3.9) in \cite{haw99}).

After this detailed variational study, which has not been 
previously developed, we are concerned with cosmological applications.

\section{Background universe and cosmological perturbations}
\label{sec-3}

From the field equations of Sec.~\ref{sec-2}, one easily finds the equations describing
a homogeneous and isotropic neutral flat universe, in which the line element is
\begin{equation}
ds^{2} = a^{2} \left[ -d\tau^{2} + dr^{2} +r^{2} d\theta^{2} 
+ r^{2} \sin^{2} \theta d\phi^{2} \right]\ .
\label{2.1} 
\end{equation}
The vector field, A, and the four-velocity, U, of the cosmic fluid have 
the covariant components $[A_{0B}(\tau ), 0, 0, 0]$ and $[-a(\tau),0,0,0]$, respectively.
The energy density is the critical one, whose present value is 
$\rho_{B0} =3H_{0}^{2}/8\pi G $, and the density of charge is $\rho_{qB}(\tau ) = 0$
for any time. In this background, Eqs.~(\ref{1.3}) reduces to 
\begin{equation}
\Xi_{B} \equiv (\nabla \cdot A)_{B} = - \frac {1}{a^{2}} [ \dot{A}_{0B} 
+ 2 \frac {\dot{a}}{a} A_{0B}] = constant   \ ,
\label{coscons}
\end{equation} 
and Eq.~(\ref{1.8}) leads to the relations
\begin{equation}
\rho^{A}_{B} = -P^{A}_{B} = -\gamma \Xi_{B}^{2} \ ,
\label{eqest}
\end{equation} 
where quantities $\rho^{A}_{B}$ and $P^{A}_{B}$ are the 
background energy density and pressure of the vector field 
$A^{\mu}$, respectively. The equation of state (\ref{eqest}) 
proves that the energy density of the background field $A^{\mu}$ play the role of 
a cosmological constant. In order to have positive values
of $\rho^{A}_{B}$ the parameter $\gamma$ must be negative.
Hereafter units are chosen in such a way that $c=8\pi G=1$; thus,
from Eqs.~(\ref{1.7}) and (\ref{1.8}) one easily get the following 
basic cosmological equation for the background evolution:
\begin{equation}
3 \frac {\dot{a}^{2}}{a^{2}}= a^{2} (\rho_{B}+\rho^{A}_{B})
\label{baseq1}
\end{equation} 
\begin{equation}
-2 \frac {\ddot{a}}{a} + \frac {\dot{a}^{2}}{a^{2}} = a^{2} (P_{B}+P^{A}_{B})
\label{baseq2}
\end{equation} 
where $\rho_{B} $ and $P_{B}$ are the background energy density and pressure 
of the cosmological fluid (baryons plus dark matter and radiation). Hereafter,
$w$ and $c_{s}^{2}$ stand for the ratios $P_{B}/\rho_{B}$ and $dP_{B}/d\rho_{B}$,
respectively.
                                                       
In next sections, perturbations are described --in the usual way-- 
with the formalism summarized in \cite{bar80} (see also \cite{huw97}).
There are three types of perturbations whose evolution is independent  during the linear 
regime. They are the so-called scalar, vector, and tensor fluctuations, which may be 
expanded in terms of the scalar, $Q^{(0)}$, vector, $Q^{(1)}_{i}$, and tensor, 
$Q^{(2)}_{ij}$, harmonics, respectively.

\subsection{Tensor perturbations}
\label{sec-3a}

There are no tensor modes involved in the 
expansion of vectors $A^{\mu} $ and $J^{\mu}$; hence, 
in the VT theory, tensor modes only appear in the expansions of the same 
quantities as in GR (metric and anisotropic part of the stress tensor) and, moreover, they 
satisfy the same equations as in GR. Therefore, we are mainly interested 
in scalar and vector modes. Metric tensor modes (gravitational waves) 
and anisotropic stress tensor components evolve as in GR, namely, 
they satisfy the equation: 
\begin{equation}
\ddot{H}_{_{T}}^{(2)} + 2 \frac {\dot{a}}{a} \dot{H}_{_{T}}^{(2)}
+k^{2} H_{_{T}}^{(2)} = P_{B} a^{2} \Pi_{_{T}}^{(2)} \ ,
\label{gw}                                      
\end{equation}   
where $\Pi_{_{T}}^{(2)} Q^{(2)}_{ij}$ 
is the tensor part of the anisotropic
stress tensor and $H_{_{T}}^{(2)} Q^{(2)}_{ij}$ the tensor part of the metric
(see \cite{bar80}).

For negligible anisotropic stress, cosmological fluctuations evolving well outside 
the effective horizon ($\dot{a}/a >> k$, see \cite{bar80,mb95})
obey the equation $\ddot{H}_{_{T}}^{(2)} + 2 (\dot{a}/a) \dot{H}_{_{T}}^{(2)}
\simeq 0$, whose general solution is $\dot{H}_{_{T}}^{(2)} \propto a^{-2} $;
hence, $\dot{H}_{_{T}}^{(2)} $ is a fast decaying mode. This means that
well after reheating, e.g. at $z=10^{8} $, 
superhorizon scales evolve in such a way that quantity
${H}_{_{T}}^{(2)} $ is almost independent 
of time. This fact will be taken into account below.

\subsection{Vector perturbations}
\label{sec-3b}                   

In the case of a flat background, the vector harmonics 
can be written as follows \cite{huw97}: $\vec{Q}^{\,\pm} = \vec{\epsilon}^{\,\pm}
\exp (i \vec{k} \cdot \vec{r})$,
where $\vec{k} $ is the wavenumber vector. 
A representation of vectors 
$\vec{\epsilon}^{\,+}$ and $\vec{\epsilon}^{\,-}$ is \cite{morsa07,morsa08}:
\begin{equation}
\epsilon^{\pm}_{1}=(\pm k_{1} k_{3} /k - i k_{2})/  \sigma \sqrt{2}\ , \,\,  
\epsilon^{\pm}_{2}=(\pm k_{2} k_{3} /k + i k_{1})/ \sigma \sqrt{2}\ , \,\,   
\epsilon^{\pm}_{3}=\mp  \sigma / k \sqrt{2} \ ,
\label{2.2}
\end{equation} 
where $\sigma = (k_{1}^{2}+k_{2}^{2})^{1/2}$.  

The first order vector part of the fields $A_{\mu}$ and $J_{\mu}$ may be written 
as follows:
\begin{equation}
A_{\mu} = [0, A^{(1)\pm} Q^{(1)\pm}_{i}] \ ,
\label{2.3}
\end{equation}   
\begin{equation}
J_{\mu} = [0, a^{2} J^{(1)\pm} Q^{(1)\pm}_{i}] \ ,
\label{2.4}
\end{equation}

As it is done in standard electromagnetism (Einstein-Maxwell theory), we define the 
covariant components of the electric and magnetic fields
($E_{\mu} $ and $B_{\mu}$) 
as follows:
\begin{equation}
E_{\mu} = F_{\mu\nu} U^{\nu} \ ,
\label{efield}
\end{equation}   
and 
\begin{equation}
B_{\mu} = \frac{1}{2} (-g)^{-1/2} \epsilon_{\mu \nu \rho \lambda} 
F^{\rho \lambda} U^{\nu} ,
\label{bfield}
\end{equation}   
where quantities $\epsilon_{\mu \nu \rho \lambda}$ are the Levy-Civita symbols. 
Then, at first order in perturbation theory, one easily finds:
\begin{equation}
F_{\mu \nu} = A_{\nu /\mu} -A_{\mu /\nu} = \left[ \begin{array} {cccc}
              0 & -aE_{1} & -aE_{2} & -aE_{3} \\
              aE_{1} & 0 & aB_{3} & -aB_{2} \\
              aE_{2} & -aB_{3} & 0 & aB_{1} \\
              aE_{3} & aB_{2} & -aB_{1} & 0
              \end{array} \right]   \ .
\label{cocom}
\end{equation}  
Equations.~(\ref{2.2})--(\ref{2.4}), and (\ref{cocom}) lead to
the following equations in momentum space:
\begin{equation}
E^{(1)\pm} = - \frac {1}{a} \dot{A}^{(1)\pm} \ , \,\,\,\, 
B^{(1)\pm} = \pm \frac {kA^{(1)\pm}}{a} \ .
\label{eb1}
\end{equation}  
Hereafter we use the following equivalences: $\vec{E} = (E_{1},E_{2},E_{3})$,
$\vec{B} = (B_{1},B_{2},B_{3})$, and $\vec{J} = (J_{1},J_{2},J_{3})$. 
Moreover, $\vec{\nabla} \cdot \vec{X}$ 
and $\vec{\nabla} \wedge \vec{X}$ stand for 
the ordinary divergence and curl of $\vec{X}$, respectively.

By using this notation and Eqs.~(\ref{eb1}) it is easily verified that,
up to first order,  
the equation 
\begin{equation}
a (\vec{\nabla} \wedge \vec{E}) + \frac {\partial}{\partial \tau} 
(a\vec{B}) = 0  \ .  
\label{posrot_b1}
\end{equation} 
is satisfied in position space. Finally, since the ordinary
divergence of any vector mode vanishes, we can write
\begin{equation}
\vec{\nabla} \cdot \vec{B} \equiv B_{i/i}= 0  \ .
\label{posdiv_b1}
\end{equation}

Let us now consider the field equations (\ref{1.3}), which may be rewritten as follows:
\begin{equation}
[\ln{(\sqrt{-g})}]_{/ \alpha} F^{\beta \alpha} +  F^{\beta \alpha}_{\,\,\,\,\,\,\, / \alpha}
= J^{\beta} -2\gamma g^{\beta \alpha} (\nabla \cdot A)_{/ \alpha} \ .
\label{fieldeq}
\end{equation}     
Since $\nabla \cdot A$ and $\nabla \cdot J $ are scalars, 
they may be expanded in scalar harmonics with no contributions
from vector modes; hence, the term $-2\gamma g^{\beta \alpha} (\nabla \cdot A)_{/ \alpha}$
vanishes in the case of vector modes and, consequently, Eqs.~(\ref{fieldeq}) reduces to 
those of Einstein-Maxwell theory. By using Eqs.~(\ref{cocom}) and (\ref{fieldeq}), it may be
easily verified that --for vector modes and up to first order, 
Einstein-Maxwell field equations and Eqs~(\ref{fieldeq}) 
may be written as follows: 
\begin{equation}
a (\vec{\nabla} \wedge \vec{B}) - \frac {\partial}{\partial \tau} 
(a\vec{E}) = a^{2}\vec{J} \ ,  
\label{posrot_e1}
\end{equation}
\begin{equation}  
\vec{\nabla} \cdot \vec{E} \equiv E_{i/i}= 0  \ .
\label{posdiv_e1}
\end{equation}

Finally, from Eqs.~(\ref{2.4}), (\ref{eb1}), and (\ref{posrot_e1}), 
the following equation describing the evolution of ${A}^{(1)\pm}$ 
is found:
\begin{equation}
\ddot{A}^{(1)\pm} + k^{2} {A}^{(1)\pm} = a^{4} {J}^{(1)\pm}      \ .  
\label{a1evol}
\end{equation}

For a given function ${J}^{(1)\pm}(\vec{k},\tau)$, 
the solution of Eq.~(\ref{a1evol}) gives ${A}^{(1)\pm}(\vec{k},\tau)$ and
then, quantities $E^{(1)\pm}$ and $B^{(1)\pm}$ are 
fixed by Eqs.~(\ref{eb1}). From these last quantities we may calculate 
$\vec{E}(\vec{x},\tau)$ and $\vec{B}(\vec{x},\tau)$ by using 
the explicit form (\ref{2.2}) of the vector harmonics. The 
resulting $\vec{E}$ and $\vec{B}$ quantities satisfy the four 
Eqs.~(\ref{posrot_b1}), (\ref{posdiv_b1}), (\ref{posrot_e1}), and 
(\ref{posdiv_e1}) in position space.   

The equations of this subsection are valid in 
the VT theory under consideration as well as in 
the standard Einstein Maxwell theory. It is due to the fact that 
vector modes do not contribute to the terms involving $\gamma$
either in Eq.~(\ref{fieldeq}) or in Eq.~(\ref{1.8}), and
these terms are responsible for all the differences between both 
theories. The predictions of these theories only may be different
due to the scalar modes involved in the 
vector fields $A^{\mu} $ and $J^{\mu} $, which are studied in next subsection.

\subsection{Scalar perturbations}
\label{sec-3c}                   

For a flat background, the 
scalar harmonics are plane waves; namely, $Q^{(0)} = \exp ({i\vec{k} \cdot \vec{r}})$.
The first order scalar contributions to vectors $A_{\mu}$ and $J_{\mu}$ are:
\begin{equation}
A_{\mu} = [\alpha^{(0)} Q^{(0)}, \beta^{(0)} Q^{(0)}_{i}] \ ,
\label{2.3b}
\end{equation}   
\begin{equation}
J_{\mu} = [0, a^{2} J^{(0)} Q^{(0)}_{i} ] \ ,
\label{2.4b}
\end{equation}
where $Q^{(0)}_{i} = (-1/k) Q^{(0)}_{/i}$.
                                                                        
Since we assume that the universe is neutral up to first order, 
the component $J_{0} $ vanishes and, moreover,
taking into account that the equation $\nabla \cdot J =0$ is not an 
equation of the VT theory, 
a scalar part $a^{2} J^{(0)} Q^{(0)}_{i}$ must be included in the expansion 
of $J_{i} $. Equations (\ref{2.3b}) and (\ref{2.4b}) are absolutely general. 

Equations (\ref{cocom}) and (\ref{2.3b}) may be combined to get:
\begin{equation}
E^{(0)} = - \frac {1}{a} (k\alpha^{(0)} +\dot{\beta}^{(0)})\ , \,\,\,\,
B^{(0)} = 0 \ .
\label{eb0}
\end{equation}    
Similarly, the scalar $\nabla \cdot A$ may be expanded in 
terms of scalar harmonics; namely, we can write:
\begin{equation}
\nabla \cdot A = \Xi_{B} (1+ \Xi^{(0)} Q^{(0)}) \ .
\label{diva_exp}
\end{equation}     
In order to calculate $\Xi^{(0)} \equiv (\nabla \cdot A)^{(0)}/ \Xi_{B}$,  
we must use the relation $\nabla \cdot A = \partial A^{\mu}/ \partial x^{\mu}
+ \Gamma^{\mu}_{\mu \nu} A^{\nu}$, which involves the Christoffeld symbols.
It is then evident that $\Xi^{(0)}$ depends on the coefficients 
$\alpha^{(0)}$ and $\beta^{(0)}$ appearing in the expansion of $A_{\mu}$
[see Eq.~(\ref{2.3b})], and also on the coefficients involved in the 
expansion of the metric components $g_{\mu \nu}$, which appear in the 
Christoffeld symbols. If the resulting $\Xi^{(0)}$ is
used to write Eqs.~(\ref{fieldeq}), (\ref{1.7}) and (\ref{1.8}) 
up to first order in scalar modes, all the equations are coupled among
them and their solution only have been found
under special assumptions (see \cite{bm092} and Sec.~\ref{intro}).
However, we have developed a method, in which we may first 
solve Eqs.~(\ref{fieldeq}) and, then, the solution can be used to 
solve Eqs.~(\ref{1.7}) and (\ref{1.8}).
Let us now describe this method and derive the evolution 
equations for the scalar modes.

Coefficients $\alpha^{(0)}$ and $\beta^{(0)}$ are not the most suitable 
ones in order to 
expand the field equations. It is preferable the use of the  
coefficients $E^{(0)}$ and $\Xi^{(0)}$,
which are gauge invariant quantities. In terms of 
these variables, Eqs.~(\ref{fieldeq}) reduce to:
\begin{equation}
2 \gamma a \Xi_{B} \dot{\Xi}^{(0)} = k E^{(0)}
\label{fe1}
\end{equation}    
\begin{equation}
\dot{E}^{(0)} = - a^{3} J^{(0)} -2\gamma k a \Xi_{B} \Xi^{(0)} -\frac {\dot{a}}{a} {E}^{(0)} \ ,
\label{fe2}
\end{equation}
and Eq.~(\ref{1.4}) may be written as follows:
\begin{equation}
\ddot{\Xi}^{(0)} + 2 \frac {\dot{a}}{a} \dot{\Xi}^{(0)} + k^{2} \Xi^{(0)}        
= -\frac {k}{2\gamma \Xi_{B}} a^{2} J^{(0)}  \ .
\label{fe12}
\end{equation}
Since Eq.~(\ref{1.4}) is a consequence of Eqs.~(\ref{fieldeq}), which are equivalent to
Eqs.~(\ref{1.3}), Eq.~(\ref{fe12}) may be easily obtained by combining Eqs.~(\ref{fe1})
and (\ref{fe2}). Hence, functions $\Xi^{(0)}$ and $E^{(0)}$ may be found
by solving Eqs.~(\ref{fe1}) and (\ref{fe2}) for a given $J^{(0)}$
plus initial values of $\Xi^{(0)}$ and $E^{(0)}$ and, then, 
the resulting $\Xi^{(0)}$ and $E^{(0)}$ functions and the chosen $J^{(0)}$  
will satisfy Eq.~(\ref{fe12}).   

Equation~(\ref{fe2}) may be easily derived from Eqs.~(\ref{fe1}) and (\ref{fe12}) and,
consequently, we may also proceed as follows: Eq.~(\ref{fe12})
is solved for a given $J^{(0)}$ and initial values of  $\Xi^{(0)}$ and     
$\dot{\Xi}^{(0)} $ and, then, the resulting solution $\Xi^{(0)}$ is
used to get function $E^{(0)}$ by using 
Eq.~(\ref{fe1}). Obviously, Eq.~(\ref{fe2}) is satisfied by the  
$\Xi^{(0)}$ and $E^{(0)}$ functions we have found by solving Eqs.~(\ref{fe12}) and
(\ref{fe1}).

It is worthwhile to point out that Eqs.~(\ref{gw}) and (\ref{fe12}) have the same form.
In fact, if we replace $H_{T}^{(2)} $ by  
$\Xi^{(0)}$ and $\Pi_{T}^{(2)} $ by $-kJ^{(0)}/2P_{B}\gamma$
in Eq.~(\ref{gw}), the resulting equation is identical to Eq.~(\ref{fe12}).
Condition $J^{(0)}=0$ in Eq.~(\ref{fe12}) is equivalent to 
condition $\Pi_{T}^{(2)} =0$ in (\ref{gw}). Hence, some previous conclusions 
about the evolution of gravitational wave modes would be also valid 
for the $\Xi^{(0)}$ evolution; in particular, for $J^{(0)} =0$ and 
superhorizon scales, the relation 
$\dot{\Xi}^{(0)} \simeq 0$ holds.

In extended electromagnetism, function $J^{(0)}$ does not vanish {\em a priori}.
Condition $J^{(0)}=0$ implies the relation $\nabla^{\mu} J_{\mu}=0$ in position 
space, but this relation is not a basic equation of the theory. By this reason, 
function $J^{(0)}$ is included in the equations derived in this section; nevertheless,
there is no --by the moment-- any physically motivated rule to build up 
this function. Anyway, for a given $J^{(0)}$ (including the cosmological 
possibility $J^{(0)}=0$), Eqs.~(\ref{fe1}) and (\ref{fe2}) involve:
the constant $\gamma < 0$, the wavenumber $k$, the scale factor, 
functions $\Xi^{(0)}$ and $E^{(0)}$, and their first order derivatives.
These equations do not involve scalar perturbations associated to the 
metric and the energy momentum tensor. They may be easily solved 
for given values of $\gamma $ and $k$, and initial values of 
$\Xi^{(0)}$ and $E^{(0)}$ (alternatively we may solve 
Eq.~(\ref{fe12}) for initial values of $\Xi^{(0)}$ and $\dot{\Xi}^{(0)}$).
The resulting function $\Xi^{(0)}$
appears in the expansion of Eqs.~(\ref{1.7}) and (\ref{1.8}) 
in scalar harmonics (see below).

The 
gauge invariant formalism described in \cite{bar80} is used in this paper; namely,
the metric, the four-velocity, and the 
part of the energy momentum tensor (\ref{1.8}) being independent of 
the parameter
$\gamma $ are all expanded as follows (in the flat case):
\begin{eqnarray}
& &
g_{00}=-a^{2}(1+2\tilde{A}Q^{(0)}), \,\,\,\, g_{0i} = -a^{2}\tilde{B}^{(0)}Q^{(0)}_{i}, \,\,\,\,
\nonumber \\
& & 
g_{ij}=a^{2}[(1+2H_{L}Q^{(0)})\delta_{ij}+2H_{T}^{(0)}Q_{ij}^{(0)}]                     
\nonumber \\
& &
U_{i} = a v^{(0)} Q^{(0)}_{i}, \,\,\,\, \rho=\rho_{B}(1+\delta Q^{(0)})
\nonumber \\
& &
T_{ij} = P_{B}(1+\pi_{L} Q^{(0)})\delta_{ij} + P_{B}\pi_{T}^{(0)}Q_{ij}^{(0)}  \ .
\end{eqnarray}   
Any other quantity as, e.g., $U_{0}$, $T_{0i}$, and so on, may be easily written in terms 
of the coefficients involved in these equations (see \cite{bar80}), which may be 
combined to build up the following gauge invariant variables:
\begin{eqnarray}
& &
\eta = (w\pi_{L} - c_{s}^{2}\delta)/w \ , \,\,\,\, v_{s}^{(0)} = v^{(0)} 
- \frac {1}{k} \dot{H}_{T}^{(0)} \ ,  
\nonumber \\
& & 
\epsilon_{m}=\delta + 3(1+w)
\frac {1}{k} \frac {\dot{a}}{a} (v^{(0)} - \tilde{B}^{(0)}) 
\nonumber \\
& &
\Phi_{A}=\tilde{A} + \frac {1}{k}  \dot{\tilde{B}}^{(0)}
+ \frac {1}{k} \frac {\dot{a}}{a} \tilde{B}^{(0)}
-\frac {1}{k^{2}}\Big(\ddot{H}_{T}^{(0)} + \frac {\dot{a}}{a} \dot{H}_{T}^{(0)}\Big)    
\nonumber \\
& &
\Phi_{H}=H_{L} + \frac {1}{3} H_{T}^{(0)} +  \frac {1}{k} \frac {\dot{a}}{a}
\tilde{B}^{(0)} - \frac {1}{k^{2}} \frac {\dot{a}}{a} \dot{H}_{T}^{(0)}
\label{invt}
\end{eqnarray}   

The complementary part of the energy momentum tensor (\ref{1.8}); namely, the 
part depending on $\gamma $ may be easily expanded in terms of scalar harmonics.
The resulting expansion involves the variable $\Xi^{(0)}$ and its first 
order time derivative (or equivalently $\Xi^{(0)}$ and $E^{(0)}$).

In order to expand Eqs.~(\ref{1.7}) and (\ref{1.8}) and the relation 
$\nabla_{\nu} T^{\mu \nu } =0$,
we use the same gauge invariant potentials and variables 
as in \cite{bar80}) (see above), plus the gauge 
invariant variables $\Xi^{(0)}$, $E^{(0)}$,
and $J^{(0)}$. The resulting equations reads as follows:
\begin{equation}
\frac {2k^{2}}{a^{2}} \Phi_{H} = \rho_{B} \epsilon_{m} 
-2\gamma \Xi_{B} \Big[  \Big( 3\frac {\dot{a}}{a^{3}}A_{0B} +
\Xi_{B} \Big) \Xi^{(0)}+\frac {A_{0B}}{a^{2}} \dot{\Xi}^{(0)} \Big]   \ ,
\label{fh} 
\end{equation}   
\begin{equation}
- \frac {k^{2}}{a^{2}} (\Phi_{A} + \Phi_{H}) = P_{B} \Pi_{T}^{(0)} \ ,
\label{fha}
\end{equation}   
\begin{equation}
\dot{v}_{s}^{(0)} + \frac {\dot{a}}{a} v_{s}^{(0)} =   
k \Phi_{A} + \frac {k}{1+w} (c_{s}^{2}\epsilon_{m} +w\eta)-
\frac {2wk}{3(1+w)} \Pi_{T}^{(0)} \ ,
\label{vs}
\end{equation}   
\begin{equation}
(\rho_{B} a^{3} \epsilon_{m})\,\dot{} = -ka^{3} (\rho_{B}+P_{B}) v_{s}^{(0)}
-2a^{2} \dot{a} P_{B} \Pi_{T}^{(0)} -ka^{3} A_{0B} J^{(0)} -3\gamma  a^{3} \Xi_{B}
A_{0B} (\rho_{B}+P_{B}) \Xi^{(0)}  \ .
\label{em} 
\end{equation}   
Eqs.~(\ref{fh}) to (\ref{em}) plus (\ref{baseq1}) and (\ref{baseq2})
may be combined to get the following equation for the evolution of 
$\epsilon_{m}$:
\begin{eqnarray}
& &
(\rho_{B} a^{3} \epsilon_{m})\,\ddot{}
+(1+3c_{s}^{2})\frac {\dot{a}}{a}(\rho_{B} a^{3} \epsilon_{m})\,\dot{}
+\Big[k^{2}c_{s}^{2} - \frac {1}{2}(\rho_{B}+P_{B})a^{2}\Big] 
(\rho_{B} a^{3} \epsilon_{m})= \nonumber \\
& &
-k^{2}(P_{B} a^{3} \eta) - 2\dot{a} (P_{B} a^{2} \Pi_{T}^{(0)})\,\dot{}
+\frac{2}{3} k^{2} (P_{B} a^{3} \Pi_{T}^{(0)})
\nonumber \\
& &                       
+2 \rho_{B} a^{2} \Big[ w - c_{s}^{2} - 
(1+c_{s}^{2})\frac {\rho^{A}_{B}} {\rho_{B}} \Big] (P_{B} a^{3} \Pi_{T}^{(0)})   
\nonumber \\
& & 
-2 \gamma a^{3} \Xi_{B} (\rho_{B}+P_{B}) (2A_{0B} \dot{\Xi}^{(0)} - 
\Xi_{B} a^{2} \Xi^{(0)})
\nonumber \\
& &
-ka^{3}\Big(A_{0B}\dot{J}^{(0)}+
\Big[(2+3c_{s}^{2})\frac {\dot{a}}{a} A_{0B}-a^{2} \Xi_{B} \Big] J^{(0)} \Big) \ .
\label{emt}     
\end{eqnarray}
For $\Xi^{(0)} =0$, $J^{(0)} =0$, and  $\rho^{A}_{B} =0$, Eqs.~(\ref{fh}) to  (\ref{emt}) 
reduce to the equations derived by Bardeen in the flat case \cite{bar80}.
For $\Xi^{(0)} =0$, $J^{(0)} =0$, and  $\rho^{A}_{B} = \rho_{\Lambda} \neq 0$, 
Eqs.~(\ref{fh}) to  (\ref{emt}) describe fluctuation evolution in 
a standard flat universe with a cosmological constant 
whose energy density is $\rho_{\Lambda}$. Finally, if $\rho^{A}_{B} = \rho_{\Lambda}$,
and the two functions  $J^{(0)} $ and $\Xi^{(0)} $ 
do not vanish at the same time, Eqs.~(\ref{fh}), (\ref{em}), and (\ref{emt}) contain new terms,
which modify the equations describing perturbation evolution in flat universes
with cosmological constant.

\section{Analyzing the basic 
differential equations of extended electromagnetism}     
\label{sec-4} 

In this section, 
it is assumed that the condition $J^{(0)} =0$ holds in cosmology,
which is equivalent to assume the well known conservation law of
Einstein-Maxwell theory ($\nabla_{\mu} J^{\mu} =0$). Under this arbitrary assumption, 
we study the background equations (\ref{coscons})--(\ref{baseq2}),
and the equations~(\ref{fe1})--(\ref{fe12}), 
and~(\ref{fh})--(\ref{emt}) describing the evolution of the first order
scalar perturbations in momentum space.

We begin with the background equations. 
Function $\rho_{B}(\tau)$ and $P_{B}(\tau)$ are given by the formulas
\begin{equation}
\rho_{B} = \rho_{Br0}(1+z)^{4}  + \rho_{Bm0}(1+z)^{3} \ ,
\, \, \, P_{B} = \rho_{Br0}(1+z)^{4}/3 \ ,
\end{equation}
where $\rho_{Br0}=8\times 10^{-34} \ gr/cm^{3}$ and 
$\rho_{Bm0}= 0.2726 \rho_{c}$ are the present energy density 
of radiation and matter 

Moreover, the baryon density is assumed to be $\rho_{Bb0}= 0.0461 \rho_{c}$,
and the value of the Hubble constant is $H_{0} = 100h \ Km s^{-1} Mpc^{-1}$ 
with $h=0.704 $. All these values are compatible with 
a certain 
version of the concordance model (see \cite{jar10}). In this model, 
the dark energy density, $\rho^{A}_{B}$, is easily obtained from the 
relation $\rho_{Bm0} + \rho_{Br0} + \rho^{A}_{B} = 3H_{0}^{2} $,
which is valid in flat backgrounds. 

Equation (\ref{baseq1}) governing  
the evolution of the scale factor may be numerically solved for
the above parameters; thus, the evolution of the scale 
factor is obtained. The resulting function $a(\tau)$ 
is necessary to study the remaining equations of the theory;
namely, Eqs.~(\ref{fe1})--(\ref{fe12})
and also Eqs.~(\ref{fh})--(\ref{emt}).

From Eq.~(\ref{eqest}) one easily gets the relation
$\Xi_{B} = \pm |\gamma|^{-1/2} (\rho^{A}_{B})^{1/2} $ (negative $\gamma $). 
Once the value of $\rho^{A}_{B} = \rho_{\Lambda}$ is
fixed (see above), this last relation leads to
$\Xi_{B} \propto S_{gn} |\gamma|^{-1/2}$,
where $S_{gn}$ only may take on the values $+1$ and $-1$.  
For a given value of 
$|\gamma|$, only the absolute value of $\Xi_{B}$ may 
be obtained (its sign is arbitrary). On account of Eq.~(\ref{coscons}),
we may also write the relation 
$A_{0B} \propto S_{gn} |\gamma|^{-1/2}$. In the 
background, quantities $|\gamma|$ and $S_{gn} $ remain
arbitrary.

Hereafter, $D^{in}$ stands for the initial
value of quantity $D$ 
whatever it may be.
Let us now consider Eqs.~(\ref{fe1})--(\ref{fe12}).
We first solve Eq.~(\ref{fe12}) by using initial values
$\Xi^{(0)in}$ and $\dot{\Xi}^{(0)in}$ at redshift $z=10^{8}$.
At this high redshift, the  
cosmological scales of interest (see below) 
are superhorizon ones; hence, taking into account the 
similarity between Eqs.~(\ref{gw}) and (\ref{fe12})
and the comments in the last paragraph 
of Sec.~\ref{sec-3a}, the condition $J^{(0)} =0$ assumed in this
section allows us to take $\dot{\Xi}^{(0)in}=0$.
Only the initial value of $\Xi^{(0)}$ may be appropriately chosen   
to integrate Eq.~(\ref{fe12}). 
The values of 
$|\gamma|$ and $S_{gn} $ are fully irrelevant to perform 
this integration.
Hereafter, numerical calculations are performed for the spatial scale 
$\tilde{L}=3 \times 10^{3} h^{-1} \ Mpc $, which reenters
the effective horizon at present time ($\tau_{0}$).
Its wavenumber is
$\tilde{k} \simeq 1.47 \times 10^{-3} $. This scale is useful 
for normalization in GR (see the Appendix).
Equation~(\ref{fe12}) has been solved for the wavenumber $\tilde{k}$ with 
the initial condition $\Xi^{(0)in}(\tilde{L}) = 10^{-4}$
in position space. 
In order to write the corresponding initial condition in momentum space, 
we use the well known relation \cite{kt94}
\begin{equation}
\left< |X(x)|^{2} \right>_{L} \simeq 
k^{3}  \left< |X(k)|^{2} \right> / 2\pi^{2} \ ,
\label{pos_mom}
\end{equation}
where $X$ is an arbitrary quantity. Evidently, this relation must
be particularized for $X=\Xi^{(0)in}$, $L=\tilde{L}$, and $k=\tilde{k} $
to calculate $\Xi^{(0)in}(\tilde{k}) $. Either this last initial quantity 
(in momentum space) or $\Xi^{(0)in}(\tilde{L}) $ (in position space) 
may be seen as a normalization 
constant. The $\Xi^{(0)}$ spectrum would be necessary to
derive the initial value of $\Xi^{(0)}(k)$ for $k \neq \tilde{k} $.

\begin{figure}[tbh]
\includegraphics[angle=0,width=0.4\textwidth]{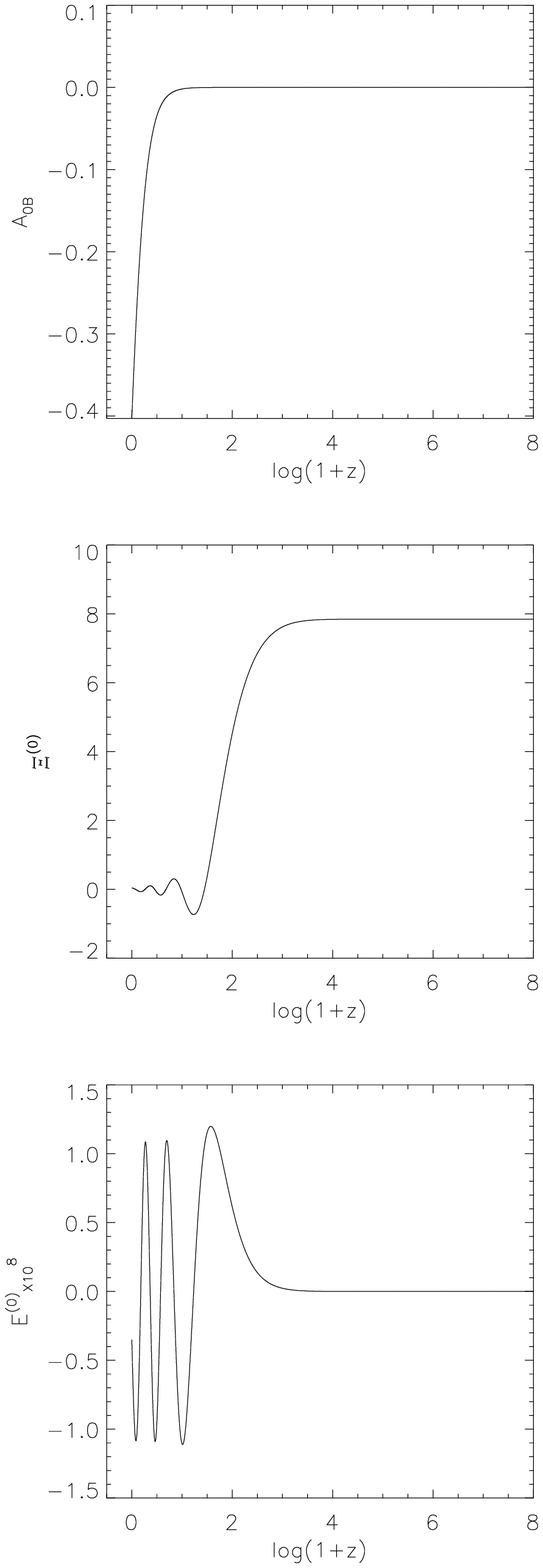}%
\caption{\label{figu1} Functions 
$A_{0B}(z)$ (top), $\Xi^{(0)}(\tilde{k},z)$ (middle) and 
$E^{(0)}(\tilde{k},z)\times 10^{8}$ (bottom)
in terms of $\log (1+z)$, for $\tilde{L} = 3000 h^{-1} \ Mpc $,
$|\gamma|=1$, $S_{gn}=+1 $, and $\Xi^{(0)in}(\tilde{L})=10^{-4}$.
}
\end{figure}

The solution of Eq.~(\ref{fe12}) plus Eq.~(\ref{fe1}) allow us
to calculate function $E^{(0)}$ which is proportional 
to $S_{gn} |\gamma|^{1/2}$. 

Numerical integrations have given the functions $A_{0B}(z)$, $\Xi^{(0)}(\tilde{k},z)$, and 
$E^{(0)}(\tilde{k},z)$ represented in Fig.~\ref{figu1}. These functions correspond to 
$|\gamma|=1$, $S_{gn}=+1 $, $k = \tilde{k} $, and $\Xi^{(0)in}(\tilde{L})=10^{-4}$.
Since the dependence of these functions in terms of the parameters 
$|\gamma|$, $S_{gn} $, and $\Xi^{(0)in}(\tilde{k})$ is known (see above), 
Fig.~\ref{figu1} contains complete information about the 
scalar modes associated to the field $A^{\mu} $ for the scale $\tilde{k} $. 
The same may be done for any linear spatial scale with the help of 
an appropriate spectrum for $\Xi^{(0)in}$.
Let us now study the
Eqs.~(\ref{fh})--(\ref{emt}) describing the evolution --in the framework of 
extended electromagnetism-- 
of the scalar modes appearing in standard GR cosmology. Equations (\ref{fh}), (\ref{em}), 
and (\ref{emt}) contain 
the terms:
\begin{equation}
\xi_{1}(\tau,k)=-2\gamma \Xi_{B} \Big[  \Big( 3\frac {\dot{a}}{a^{3}}A_{0B} +
\Xi_{B} \Big) \Xi^{(0)}+\frac {A_{0B}}{a^{2}} \dot{\Xi}^{(0)} \Big]   \ ,
\label{fhnew} 
\end{equation}
\begin{equation}
\xi_{2}(\tau,k)=-3\gamma a^{3} \Xi_{B}
A_{0B} (\rho_{B}+P_{B}) \Xi^{(0)}  \ ,
\label{emnew}
\end{equation}  
\begin{equation}  
\xi_{3}(\tau,k) = -2 \gamma a^{3} \Xi_{B} (\rho_{B}+P_{B}) (2A_{0B} \dot{\Xi}^{(0)} - 
\Xi_{B} a^{2} \Xi^{(0)}) \ ,
\label{emtnew}      
\end{equation}  
respectively. These terms --which appear in 
extended electromagnetism but not 
in Einstein theory with cosmological constant-- may be 
calculated, for the scale $\tilde{k} $, by using the integration 
data used to build up Fig.~\ref{figu1} and, then, these terms may be compared
with appropriate terms involved in 
GR equations (for the same wavenumber).

Taking into account that $\Xi_{B}$ and $A_{0B}$ are proportional to
$S_{gn} |\gamma|^{-1/2}$ and also that $\Xi^{(0)}$ does not depend
on $S_{gn}$ and $|\gamma|$, it is trivially proved that 
quantities $\xi_{1}$, $\xi_{2}$, and $\xi_{3}$ are also independent 
of $S_{gn}$ and $|\gamma|$. For the wavenumber $\tilde{k}$, these quantities are proportional to  
the number $\Xi^{(0)in}(\tilde{L})$.

As it follows from Eqs.~(\ref{fh}), (\ref{em}) (\ref{emt}),
quantities $\xi_{1}$, $\xi_{2}$, and $\xi_{3}$ are to be
compared with the GR values of the terms 
\begin{equation}  
\Upsilon_{1}(\tau,k)=\rho_{B} \epsilon_{m}  \ ,
\label{fhgr}      
\end{equation}  
\begin{equation}  
\Upsilon_{2}(\tau,k)=-ka^{3} (\rho_{B}+P_{B}) v_{s}^{(0)}  \ ,
\label{emgr}      
\end{equation}  
\begin{equation}  
\Upsilon_{3}(\tau,k)=\Big[ \frac {1}{2}(\rho_{B}+P_{B})a^{2} - k^{2}c_{s}^{2} \Big] 
(\rho_{B} a^{3} \epsilon_{m})  \ ,
\label{emtgr}      
\end{equation} 
respectively. After the estimation of $\Upsilon_{1}$, $\Upsilon_{2}$, and 
$\Upsilon_{3}$ in standard cosmology (based on GR), the three functions
$r_{i}(\tau,k)=|\xi_{i}(\tau,k)/\Upsilon_{i}(\tau,k)|$ may be calculated. Evidently,
for very small $r_{i} $ values, GR and extended electromagnetism
would lead to the same differential equations for the evolution 
of the GR scalar perturbations,
whereas $r_{i} $ values of the order of $10^{-3} $ or greater
would suggest relevant differences with respect to GR.
If differences are expected, the matter power spectrum $P(k) $ and
the angular power spectra of the CMB should be accurately estimated 
by using numerical codes 
as CMBFAST \cite{seza96} and CAMB \cite{lew00}. 
These accurate calculations -under general enough initial conditions-- are beyond the 
scope of this paper. 

The $r_{i}$ ratios will be estimated for the wavenumber 
$\tilde{k} $. The corresponding spatial scale is useful 
for normalization in GR, which is necessary to estimate the 
functions $\Upsilon_{i}(\tau,\tilde{k}) $. The method used for
the estimation of these functions and for normalization in standard 
GR cosmology are described in the Appendix.   

\begin{figure}[tbh]
\includegraphics[angle=0,width=0.8\textwidth]{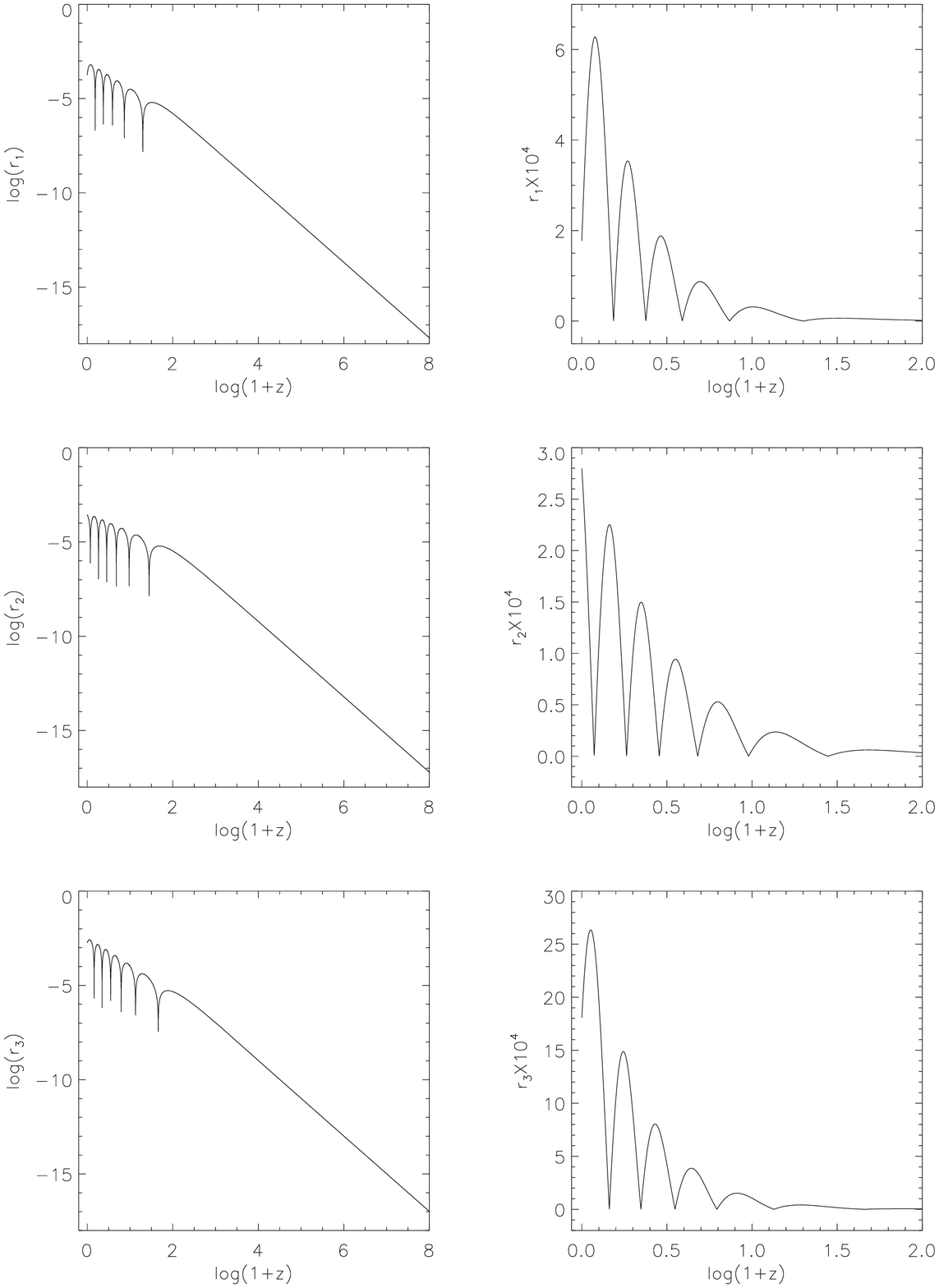}%
\caption{\label{figu2} 
Left: functions 
$\log[r_{1}(\tilde{k},z)]$ (top), $\log[r_{2}(\tilde{k},z)]$ (middle) and 
$\log[r_{3}(\tilde{k},z)]$ (bottom)
in terms of $\log (1+z)$, from $z=10^{8} $ to $z=0$. The wavenumber 
$\tilde{k} $ is the same as in Fig.~\ref{figu1}. Right: functions 
$r_{1}(\tilde{k},z) \times 10^{4}$ (top), $r_{2}(\tilde{k},z) \times 10^{4}$ (middle) and 
$r_{3}(\tilde{k},z) \times 10^{4}$ (bottom) in terms of $\log (1+z)$, 
from $z=10^{2} $ to $z=0$, for the same wavenumber as in the                  
left panels, 
}
\end{figure}

The three functions $r_{i}(z,\tilde{k})$ are represented 
in Fig.~\ref{figu2} for $\Xi^{(0)in}(\tilde{L})=10^{-4}$. Two panels (left and right) 
show the evolution of each ratio $r_{i} $. The evolutions of the 
three ratios are similar. From $z=10^{8} $ to $z \sim 10^{2} $,
the chosen spatial scale is well outside the effective horizon 
and all the ratios increase without oscillations (see left panels);
however, for $z < 10^{2} $, there are oscillations whose 
amplitudes grow as $z $ decreases (see right panels). It is due to the fact that 
our spatial scale and that 
of the effective horizon come near as the redshift decreases
(they have been chosen to be identical at $z=0 $). The maximum values of 
the ratios are $r_{1} \simeq 6.3 \times 10^{-4} $, 
$r_{2} \simeq  2.8 \times 10^{-4}  $, and 
$r_{3} \simeq  2.6 \times 10^{-3} $. These maximum values 
are all reached close to $z=0$. They are small enough to ensure that,
for $\Xi^{(0)in}(\tilde{L})=10^{-4}$, the chosen scale evolves as in GR.

\begin{figure}[tbh]
\includegraphics[angle=0,width=0.4\textwidth]{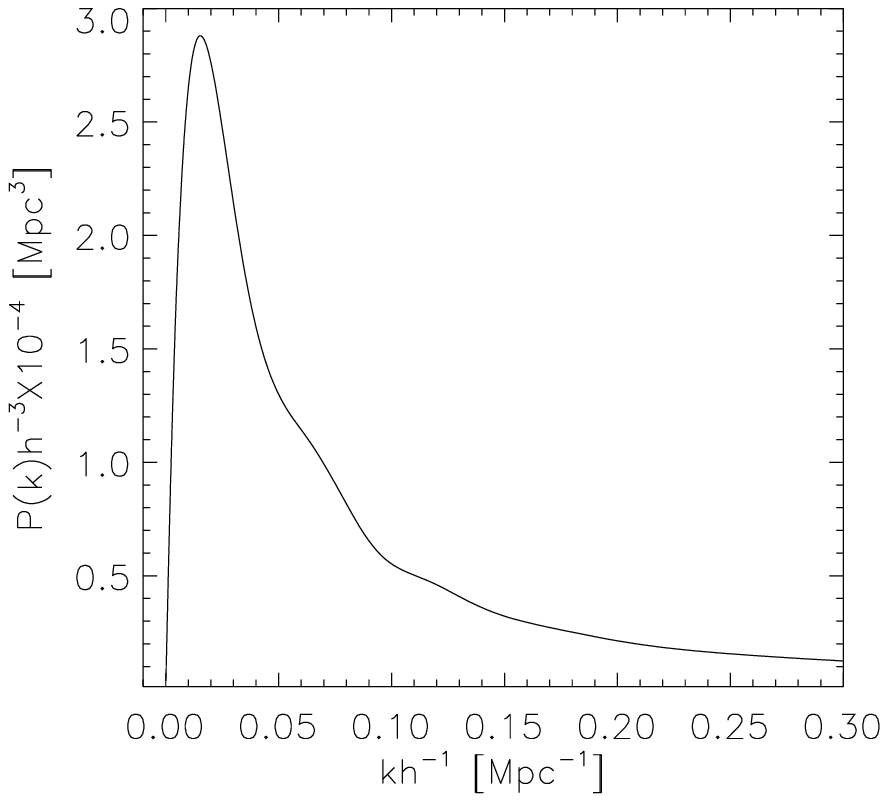}%
\caption{\label{figu3} matter power spectrum --estimated with CMBFAST- for 
the chosen version of the concordance model.
}
\end{figure}

\section{Discussion and conclusions}     
\label{sec-5} 

An exhaustive variational formulation of extended electromagnetism 
has been presented in Sec.~\ref{sec-2}. In particular, 
the energy momentum tensor of the vector field $A^{\mu} $ has been found.
This tensor has been analyzed to conclude that 
the last term of Eq.~(\ref{1.8})
has not the same sign as in previous calculations \cite{bm092}. 
Our sign appears as a result of the relation 
$\delta_{_g} (\sqrt {-g} J^{\alpha}) = -\delta_{_g} (\sqrt {-g} J^{^{A} \alpha}) $,
which is consistent with the conservation law of $J_{\mu} + J^{A}_{\mu} $.
This sign is only compatible with a negative constant $\gamma $ (see Sec.\ref{sec-3}). 
Our formulation also leads to Eqs.~(\ref{1.6}) giving the components of the Lorentz force
and to Eq.~(\ref{1.9}).

We have emphasized that,
for vector and tensor linear perturbation of the Friedman-Robertson-Walker universe,
extended electromagnetism
is fully equivalent to Einstein-Maxwell theory.
Although this fact was already known (see \cite{bm092}), our 
detailed but brief study of vector and tensor perturbations 
is necessary to describe our approach to the evolution
of scalar modes and, moreover, this study exhibits novel aspects. 
Note that: (i) our brief considerations about tensor perturbations
have been used to study the evolution of some scalar modes [analogy between 
Eqs.~(\ref{gw}) and (\ref{fe12})], (ii)
our analysis of vector perturbations leads to new relations in momentum space 
derived from Eqs.~(\ref{2.2}) and, (iii) this analysis suggests
the definition of the gauge invariant 
quantity $E^{(0)}$, which plays a crucial role in our description 
of the scalar modes.

A new approach to deal with the evolution of the scalar modes 
in extended electromagnetism has been described. It is gauge invariant 
and fully general. This formalism leads to 
Eqs~(\ref{fe1})--(\ref{fe12}) and Eqs.~(\ref{fh})--(\ref{emt}), which 
are the most important findings of this paper.

The scalar modes associated to the vector field $A^{\mu} $ 
are assumed to be $\Xi^{(0)} $ and $E^{(0)} $. Then, in Sec.~\ref{sec-3c},
it is proved that the evolution equations of these gauge 
invariant modes [Eqs~(\ref{fe1})--(\ref{fe12})] do not involve any other scalar mode
(excepting $J^{(0)}$). These simple equations may be easily solved by using standard numerical methods
for any given function $J^{(0)}$.

In Sec.~\ref{sec-4}, where the simplifying condition $J^{(0)}=0$ is assumed,
the solution of Eqs~(\ref{fe1})--(\ref{fe12}) may be found by using
the initial value of $\Xi^{(0)} $ at $z \sim 10^{8}$, whereas the initial values of
both $\dot{\Xi}^{(0)} $ and $E^{0} $ [related by Eq.~(\ref{fe1})]
may be neglected. The solution corresponding 
to a particular wavenumber $\tilde{k} $ (reentering the effective horizon 
at present time) is presented in Fig.~\ref{figu1}.
There is no problem to integrate Eqs~(\ref{fe1})--(\ref{fe12}) for any other
scale. Since we have the numerical solutions of these equations, 
the correction terms $\xi_{i} $
appearing in Eqs.~(\ref{fh})--(\ref{emt}) may be numerically treated as known functions
of $k$ and $\tau $ and, consequently,
these last equations only involve --as unknown functions-- the scalar modes appearing 
in GR. Their numerical solution 
has not been found in this paper; where the correction terms $\xi_{i} $ have been
compared with appropriate terms of the GR equations ($\Upsilon_{i} $) for suitable
values of $k$ and $\Xi^{(0)in} $. For the chosen scale $\tilde{k} $ and $\Xi^{(0)in} (\tilde{L}) <10^{-4}$,
the scalar modes would evolve as in GR. The spectrum of $\Xi^{(0)} $ and 
the standard power spectrum $P(k)$,
at initial time, would be necessary to perform similar 
comparisons for all the cosmological scales. 
If all the scales are found to evolve as in GR, both theories 
are equivalent from the cosmological point of view; 
on the contrary, some appropriate code, as e.g., CMBFAST or CAMB, 
may be modified to estimate --in the VT-- the angular power spectrum of the CMB, 
the matter power spectrum, and so on.

In the background, functions $A_{0B}(\tau) $ and $\Xi_{B}(\tau) $ 
cannot be fully fixed.
Both functions are proportional to $S_{gn} |\gamma |^{-1/2}$,
but these parameters are arbitrary. Moreover,
function $\Xi^{(0)} $ and
the correction terms $\xi_{i} $ defined in Eqs.~(\ref{fhnew})--(\ref{emtnew})
are independent of parameters $S_{gn} $
and $|\gamma |$. Hence, cosmological considerations 
cannot fix the values of these parameters. 
It is not surprising, since it is well known (see \cite{wil06}) that  
the theories based on the 
Lagrangian (\ref{1.2}) with $J^{\mu} =0$ (no currents) cannot completely fix the vector field.
 
Condition $J^{(0)} =0$ has been assumed to be valid in cosmology;
nevertheless, this condition is not strictly required by 
extended electromagnetism. The question is: What would be a scalar $J^{(0)} $-current
in cosmology? More research about this scalar mode and its meaning is 
being carried out. 

Finally, let us discuss in detail the fact that our energy 
momentum tensor and that found in \cite{bm092} have opposite signs. 
The possible consequences of this difference deserve special attention.

If the relation $\gamma = 2\xi $ is satisfied, 
our Lagrangian (with $\gamma $) is identical to that used in \cite{bm092}
(including $\xi $). In spite of this fact, opposite signs appear in 
the energy momentum tensors.
As it is explained in Sec.~\ref{sec-2} and summarized in the first 
paragraph of this section, our sign is obtained --from the common Lagrangian-- 
with right 
variational calculations based on the true conservation law of the 
theory. 
Equation (4) is actually the conservation law 
satisfied in extended electromagnetism; 
however, if we take $\nabla^{\mu} J_{\mu} =0$,
which is not an equation of the theory, but the conservation 
law of standard electromagnetism, 
the opposite sign is easily found 
in the resulting energy momentum tensor. This sign is not right. 

For an arbitrary positive $\xi $ value
and for the corresponding negative value $\gamma = -\xi/2 $,
the energy-momentum tensor in \cite{bm092} is identical to our 
energy-momentum tensor. Hence, 
the Einstein equations are also indistinguishable for these values of
$\xi $ (positive) and $\gamma $ (negative). However, the field equations of the 
vector field $A^{\mu} $ are
different for the same values; namely, for $\gamma = -\xi/2 $.

Equations~(\ref{1.3}) may be written in the form
$\nabla^{\nu} F_{\mu \nu} = J_{\mu} -2 \gamma \nabla_{\mu} (\nabla \cdot A)$,  
where $\nabla \cdot A = \nabla_{\mu} A^{\mu} $, and these field equations 
are to be compared with the 
equations 
$\nabla^{\nu} F_{\mu \nu} = J_{\mu} - \xi \nabla_{\mu} (\nabla \cdot A)$ 
appearing in \cite{bm092}. This comparison shows that
the terms $-2 \gamma \nabla_{\mu} (\nabla \cdot A)$ and
$- \xi \nabla_{\mu} (\nabla \cdot A)$ --which modify the field equations of 
standard electromagnetis-- have opposite signs for $\gamma = -\xi/2 $.
Since the resulting $A^{\mu} $ field equations are different, distinct predictions
seem to be unavoidable in  general 
nonlinear applications of extended electromagnetism. The discussion 
of these nonlinear cases is beyong the scope of this paper, where 
we are concerned with cosmological linear applications of the theory.

Actually, both signs lead to the same conclusions 
for first order perturbations of Minkowski and Robertson-Walker space-times, namely,
for the cases considered, e.g., in \cite{bm009} and \cite{bm010}.
It is due to the fact that, according to Eqs.~(\ref{coscons}), (\ref{eqest}), 
and~(\ref{fe12}), the signs of $\Xi_{B}$ and $\Xi^{(0)} \equiv (\nabla \cdot A)^{(0)}/ \Xi_{B}$ 
are arbitrary and,
consequently, for $\gamma = -\xi/2 $, these signs may be chosen 
to make identical the scalar parts of the terms $-2 \gamma \nabla_{\mu} (\nabla \cdot A)$ and
$- \xi \nabla_{\mu} (\nabla \cdot A)$. Thus, the linearized 
$A^{\mu} $ field equations derived in our paper become 
equivalent to those of previous papers \citep{bm092, bm009, bm010}.
Since the energy momentum tensors are also identical for $\gamma = -\xi/2 $, 
the modes of positive energy coincide, and the conclusions of papers
\cite{bm009} and \cite{bm010} (with $\xi >0$) may be also obtained 
here for $\gamma < 0$.
However, only our signs are right and, in general, only our $A^{\mu} $ field 
equations should be applied in nonlinear cases.

\begin{acknowledgments}
This work has been supported by the Spanish
Ministerio de Ciencia e Innovaci\'on, MICINN-FEDER project
FIS2009-07705. We thank J.A. Morales-LLadosa
for useful discussion.
\end{acknowledgments}

\appendix*

\section{Estimating the $\Upsilon_{i}$ functions in GR}

Since WMAP observations strongly suggest that 
cosmological perturbations are adiabatic, only the 
case $\eta =0$ is considered in this section.

For cosmological perturbations evolving outside 
the effective horizon ($k<aH/2\pi$), the evolution is essentially
independent of the microphysics. 
This means that the anisotropic stress due
to neutrinos may be neglected ($\Pi_{T}^{(0)}=0$),
and also that, 
in spite of the tight coupling between photons and baryons
(see \cite{mb95}) at $z > 1100 $, 
the transfer of energy and momentum 
between these two species may be forgotten and, consequently,
the corresponding fluids may be treated as independent.
This means that, for superhorizon scales,
a good enough estimate of functions $\Upsilon_{i}$ may be done
by solving Eqs.~(\ref{fh})--(\ref{emt}) for $\Pi_{T}^{(0)}=\eta=0$.
In this case, 
Eqs.~(\ref{fh})--(\ref{emt}) lead to:
\begin{equation}
\ddot{\Psi}_{m}
+(1+3c_{s}^{2})\frac {\dot{a}}{a}\dot{\Psi}_{m}
+\Big[k^{2}c_{s}^{2} - \frac {1}{2}(\rho_{B}+P_{B})a^{2}\Big] 
\Psi_{m}=0 \ ,
\label{emt_app}
\end{equation}
\begin{equation}
\dot{\Psi}_{m} = -ka^{3} (\rho_{B}+P_{B}) v_{s}^{(0)}
\ ,
\label{em_app} 
\end{equation}
\begin{equation}
\dot{v}_{s}^{(0)} + \frac {\dot{a}}{a} v_{s}^{(0)} =  \frac {1}{a} 
\Big[\frac {kc_{s}^{2}}{(\rho_{B}+P_{B})a^{2}}
-\frac{1}{2k}\Big]\Psi_{m} \ ,
\label{vs_app}
\end{equation}
\begin{equation}
\Phi_{H} = \frac {\Psi_{m}}{2ak^{2}} \ ,
\label{ph_app}
\end{equation}
where $\Psi_{m}= \rho_{B} a^{3} \epsilon_{m} $. This 
system of equations must be solved together with 
the background equations for appropriate initial
conditions at $z=10^{8} $. 

The background is a flat universe with cosmological
constant. The energy densities of matter, radiation and 
vacuum correspond to the concordance model (see above). 
The background differential 
equations may be easily integrated to get $a(\tau)$,
$\rho_{B}(\tau)$, and $P_{B}(\tau)$. 

The integration of Eqs.~(\ref{emt_app})--(\ref{vs_app})  
only requires $\epsilon_{m}^{in}$ and
$v_{s}^{(0)in}$ at $z=10^{8}$. In fact, from the first of these values 
one easily obtains $\Psi_{m}^{in}$, and 
the initial value of $\dot{\Psi}_{m} $
may be then obtained by substituting $v_{s}^{(0)in}$
into Eq.~(\ref{vs_app}). The second order differential 
equation (\ref{emt_app}) may be integrated by using 
$\Psi_{m}^{in}$ and $\dot{\Psi}_{m}^{in}$. 
Function $\Psi_{m}(\tau)$ is then known and 
$v_{s}^{(0)in}$ may be used to solve Eq.~(\ref{vs_app})
and get function $v_{s}^{(0)}(\tau)$.

Since $\epsilon_{m}$ and $v_{s}^{(0)}$ are gauge invariant 
quantities, their initial values may be calculated in any gauge.
We have used the synchronous gauge to perform this calculation.
For superhorizon scales, equation~(96) of reference 
\cite{mb95} may be used to easily get the following initial conditions:
\begin{eqnarray}
& &
\delta_{\gamma}^{in} = -\frac{2}{3} C (k \tau^{in} )^{2}, \,\,\,\, 
\delta_{c}^{in} = \delta_{b}^{in} =
\frac{3}{4} \delta_{\gamma}^{in}, \,\,\,\,
\nonumber \\
& & 
v^{(0)in}_{c}=0, \,\,\,\, v^{(0)in}_{\gamma}=v^{(0)in}_{b}=-\frac{1}{18} C k^{3} (\tau^{in})^{3}, 
\,\,\,\,        
\nonumber \\
& &
H_{L}^{in}=\frac {1}{6} C (k\tau^{in} )^{2}, \,\,\,\, H_{T}^{(0)^{in}}=-6C(1+\frac{1}{18})(k\tau^{in} )^{2},
\label{init}
\end{eqnarray}   
where the conformal time $\tau^{in} $ is that corresponding to the chosen 
initial redshift $z=10^{8} $, $C$ is a normalization constant, and 
the subscripts $\gamma $, $b$, and $c $ stand for photons, baryons, and 
cold dark matter, respectively. The fluid formed by these three components has 
the following density contrast and peculiar velocity \cite{mb95} :
\begin{eqnarray}
& &
\delta = (\rho_{Bb} \delta_{b }+ \rho_{Bc} \delta_{c }+\rho_{B\gamma} \delta_{\gamma }) / 
\rho_{B}
\nonumber \\
& & 
v^{(0)}  =  [(\rho_{Bb}+P_{Bb}) v^{(0)}_{b }+ (\rho_{Bc}+P_{Bc}) v^{(0)}_{c }
+(\rho_{B\gamma}+P_{B\gamma}) v^{(0)}_{\gamma })]/(\rho_{B}+P_{B})  \ .
\label{superp}
\end{eqnarray}  
The initial values of              
$\epsilon_{m} $ and $v_{s}^{(0)}$ may be easily calculated taking into account 
Eqs.~(\ref{invt}), (\ref{init}), and (\ref{superp}). 
Equations.~(\ref{emt_app})--(\ref{vs_app}) may be then solved.

The estimation of quantities $\Upsilon_{i}$ requires normalization. The question is:
How can we find a good enough value of the normalization constant $C$? 
It is well known that, for superhorizon scales, the gauge invariant quantity
\begin{equation}
\zeta=\frac{2}{3} \,\, \frac{\Phi_{H} +(aH)^{-1} \dot{\Phi}_{H} }{1+w}
+\Phi_{H}\Big[ 1+\frac{2}{9} \Big(\frac {k} {aH} \Big)^{2} \frac {1}{1+w} \Big]
\label{fseta}
\end{equation}
is conserved \cite{bran85} and, moreover, at horizon crossing, the relation 
\begin{equation}
\delta (k,\tau) = O(1) \frac {\zeta} {1+w}
\label{pre_cont}
\end{equation}
is satisfied, where $O(1) $ is a number of order unity
(see \cite{bran85} and references cited therein). Then, normalization may be achieved as 
follows: in a first step, 
the matter power spectrum at present time $P(k,\tau_{0})$
is obtained, by using CMBFAST, for a certain version of the concordance 
model (see \cite{da11} for details).
The resulting  spectrum is represented 
in Fig.~\ref{figu3}. From it, we can estimate
$P(\tilde{k},\tau_{0})$. In a second step,       
Eqs.~(\ref{emt_app})--(\ref{vs_app}) are numerically
solved for the scale $\tilde{k} $
and for an arbitrary $C$ value and, then, by combining
Eqs.~(\ref{em_app}), (\ref{ph_app}), and (\ref{fseta}), 
the function $\zeta(\tilde{k},\tau)$ may be easily calculated. 
Finally, in a last step,
the value of the normalization constant $C$ is fixed.
It is done by using Eq.~(\ref{pre_cont}) to calculate $\delta (\tilde{k},\tau_{0})$,
and taken into account that
the resulting $\delta (\tilde{k},\tau_{0})$ quantity must be identical to
$P^{1/2}(\tilde{k},\tau_{0})$ for the right $C$ value. 

The $\delta (\tilde{k},\tau_{0})$ value obtained 
from the spectrum of Fig.~\ref{figu3} and Eq.~(\ref{pos_mom}) may be easily used 
to estimate the contrast $\delta (\tilde{L},\tau_{0})$ in position space, the resulting 
value is close to $10^{-3} $ as it is expected for this scale.

\end{document}